\documentclass[twocolumn]{aastex61}
\pdfoutput=1
\usepackage{graphicx}
\usepackage{amsmath}
\usepackage{hyperref}

\shorttitle{ASKAP detection of FRB~170107}
\shortauthors{Bannister et al.}

\begin{document}

\title{The detection of an extremely bright fast radio burst in a phased array feed survey}

\correspondingauthor{Keith Bannister}
\email{keith.bannister@csiro.au}

\author{K.~W. Bannister}
\affiliation{Australia Telescope National Facility, CSIRO Astronomy and Space Science, PO Box 76, Epping, NSW 1710, Australia}

\author{R.~M. Shannon}
\affiliation{International Centre for Radio Astronomy Research, Curtin University, Bentley, WA 6102, Australia}
\affiliation{Australia Telescope National Facility, CSIRO Astronomy and Space Science, PO Box 76, Epping, NSW 1710, Australia}

\author{J.-P. Macquart}
\affiliation{International Centre for Radio Astronomy Research, Curtin University, Bentley, WA 6102, Australia}
\affiliation{ARC Centre of Excellence for All-sky Astrophysics (CAASTRO), Australia}

\author{C. Flynn}
\affiliation{Swinburne University of Technology, PO Box 218, Hawthorn, VIC 3122, Australia}
\affiliation{ARC Centre of Excellence for All-sky Astrophysics (CAASTRO), Australia}

%(CRAFT)

\author{P.~G. Edwards}
\affiliation{Australia Telescope National Facility, CSIRO Astronomy and Space Science, PO Box 76, Epping, NSW 1710, Australia}

\author{M. O'Neill}
\affiliation{Swinburne University of Technology, PO Box 218, Hawthorn, VIC 3122, Australia}
\affiliation{ARC Centre of Excellence for All-sky Astrophysics (CAASTRO), Australia}

\author{S. Os{\l}owski}
\affiliation{Swinburne University of Technology, PO Box 218, Hawthorn, VIC 3122, Australia}

\author{M. Bailes}
\affiliation{Swinburne University of Technology, PO Box 218, Hawthorn, VIC 3122, Australia}

\author{B. Zackay}
\affiliation{Weizmann Institute of Science, Herzl St 234, Rehovot, Israel}

\author{N. Clarke}
\affiliation{International Centre for Radio Astronomy Research, Curtin University, Bentley, WA 6102, Australia}

\author{L.~R. D'Addario}
\affiliation{Cahill Center for Astronomy and Astrophysics, California Institute of Technology, MC 249-17, Pasadena, CA 91125, USA}

\author{R. Dodson}
\affiliation{ICRAR, The University of Western Australia, 35 Stirling Highway, Crawley WA 6009, Australia}

\author{P.~J. Hall}
\affiliation{International Centre for Radio Astronomy Research, Curtin University, Bentley, WA 6102, Australia}

\author{A. Jameson}
\affiliation{Swinburne University of Technology, PO Box 218, Hawthorn, VIC 3122, Australia}

\author{D. Jones}
\affiliation{Space Science Institute, Boulder, CO 80301, USA}

\author{R. Navarro}
\affiliation{Jet Propulsion Laboratory, California Institute of Technology, Pasadena, California, USA.}

\author{J.~T. Trinh}
\affiliation{US Navy, P.O. Box 5000, Corona, CA 92878, USA}

%(CRAFT stage-1)

\author{J. Allison}
\affiliation{Australia Telescope National Facility, CSIRO Astronomy and Space Science, PO Box 76, Epping, NSW 1710, Australia}

\author{C.~S. Anderson}
\affiliation{CSIRO Astronomy and Space Science, 26 Dick Perry Avenue, Kensington WA 6151, Australia}

\author{M. Bell}
\affiliation{Australia Telescope National Facility, CSIRO Astronomy and Space Science, PO Box 76, Epping, NSW 1710, Australia}

\author{A.~P. Chippendale}
\affiliation{Australia Telescope National Facility, CSIRO Astronomy and Space Science, PO Box 76, Epping, NSW 1710, Australia}

\author{J.~D. Collier}
\affiliation{Australia Telescope National Facility, CSIRO Astronomy and Space Science, PO Box 76, Epping, NSW 1710, Australia}
\affiliation{Western Sydney University, Locked Bag 1797, Penrith South, NSW 1797, Australia}

\author{G. Heald}
\affiliation{CSIRO Astronomy and Space Science, 26 Dick Perry Avenue, Kensington WA 6151, Australia}

\author{I. Heywood}
\affiliation{Australia Telescope National Facility, CSIRO Astronomy and Space Science, PO Box 76, Epping, NSW 1710, Australia}
\affiliation{Department of Physics and Electronics, Rhodes University, PO Box 94, Grahamstown, 6140, South Africa}

\author{A.~W. Hotan}
\affiliation{Australia Telescope National Facility, CSIRO Astronomy and Space Science, PO Box 76, Epping, NSW 1710, Australia}

\author{K. Lee-Waddell}
\affiliation{Australia Telescope National Facility, CSIRO Astronomy and Space Science, PO Box 76, Epping, NSW 1710, Australia}

\author{J.~P. Madrid}
\affiliation{Australia Telescope National Facility, CSIRO Astronomy and Space Science, PO Box 76, Epping, NSW 1710, Australia}

\author{J. Marvil}
\affiliation{Australia Telescope National Facility, CSIRO Astronomy and Space Science, PO Box 76, Epping, NSW 1710, Australia}

\author{D. McConnell}
\affiliation{Australia Telescope National Facility, CSIRO Astronomy and Space Science, PO Box 76, Epping, NSW 1710, Australia}

\author{A. Popping}
\affiliation{ICRAR, The University of Western Australia, 35 Stirling Highway, Crawley WA 6009, Australia}
\affiliation{ARC Centre of Excellence for All-sky Astrophysics (CAASTRO), Australia}

\author{M.~A. Voronkov}
\affiliation{Australia Telescope National Facility, CSIRO Astronomy and Space Science, PO Box 76, Epping, NSW 1710, Australia}

\author{M.~T. Whiting}
\affiliation{Australia Telescope National Facility, CSIRO Astronomy and Space Science, PO Box 76, Epping, NSW 1710, Australia}

%(ACES)

\author{G.~R. Allen}
\affiliation{Australia Telescope National Facility, CSIRO Astronomy and Space Science, PO Box 76, Epping, NSW 1710, Australia}

\author{D.~C.-J. Bock}
\affiliation{Australia Telescope National Facility, CSIRO Astronomy and Space Science, PO Box 76, Epping, NSW 1710, Australia}

\author{D.~P. Brodrick}
\affiliation{Australia Telescope National Facility, CSIRO Astronomy and Space Science, PO Box 76, Epping, NSW 1710, Australia}

\author{F. Cooray}
\affiliation{1-7 Rowe Street, Eastwood, NSW 2122, Australia}

\author{D.~R. DeBoer}
\affiliation{Department of Astronomy, University of California, Berkeley, CA, USA}

\author{P.~J. Diamond}
\affiliation{SKA Organisation, Jodrell Bank, Lower Withington, Cheshire, SK11 9DL, UK}

\author{R. Ekers}
\affiliation{Australia Telescope National Facility, CSIRO Astronomy and Space Science, PO Box 76, Epping, NSW 1710, Australia}

\author{R.~G. Gough}
\affiliation{Australia Telescope National Facility, CSIRO Astronomy and Space Science, PO Box 76, Epping, NSW 1710, Australia}

\author{G.~A. Hampson}
\affiliation{Australia Telescope National Facility, CSIRO Astronomy and Space Science, PO Box 76, Epping, NSW 1710, Australia}

\author{L. Harvey-Smith}
\affiliation{Australia Telescope National Facility, CSIRO Astronomy and Space Science, PO Box 76, Epping, NSW 1710, Australia}

\author{S.~G. Hay}
\affiliation{CSIRO Data 61,Corner Vimiera \& Pembroke Roads, Marsfield NSW 2122, Australia}

\author{D.~B. Hayman}
\affiliation{Australia Telescope National Facility, CSIRO Astronomy and Space Science, PO Box 76, Epping, NSW 1710, Australia}

\author{C.~A. Jackson}
\affiliation{International Centre for Radio Astronomy Research, Curtin University, Bentley, WA 6102, Australia}

\author{S. Johnston}
\affiliation{Australia Telescope National Facility, CSIRO Astronomy and Space Science, PO Box 76, Epping, NSW 1710, Australia}

\author{B.~S. Koribalski}
\affiliation{Australia Telescope National Facility, CSIRO Astronomy and Space Science, PO Box 76, Epping, NSW 1710, Australia}

\author{N.~M. McClure-Griffiths}
\affiliation{Research School of Astronomy \& Astrophysics, Australian National University, Canberra ACT 2611 Australia}

\author{P. Mirtschin}
\affiliation{Australia Telescope National Facility, CSIRO Astronomy and Space Science, PO Box 76, Epping, NSW 1710, Australia}

\author{A. Ng}
\affiliation{Australia Telescope National Facility, CSIRO Astronomy and Space Science, PO Box 76, Epping, NSW 1710, Australia}

\author{R.~P.~ Norris}
\affiliation{Australia Telescope National Facility, CSIRO Astronomy and Space Science, PO Box 76, Epping, NSW 1710, Australia}
\affiliation{Western Sydney University, Locked Bag 1797, Penrith South, NSW 1797}

\author{S.~E. Pearce}
\affiliation{Australia Telescope National Facility, CSIRO Astronomy and Space Science, PO Box 76, Epping, NSW 1710, Australia}

\author{C.~J. Phillips}
\affiliation{Australia Telescope National Facility, CSIRO Astronomy and Space Science, PO Box 76, Epping, NSW 1710, Australia}

\author{D.~N. Roxby}
\affiliation{Australia Telescope National Facility, CSIRO Astronomy and Space Science, PO Box 76, Epping, NSW 1710, Australia}

\author{E.~R. Troup}
\affiliation{Australia Telescope National Facility, CSIRO Astronomy and Space Science, PO Box 76, Epping, NSW 1710, Australia}

\author{T. Westmeier}
\affiliation{ICRAR, The University of Western Australia, 35 Stirling Highway, Crawley WA 6009, Australia}

\begin{abstract}
We report the detection of an ultra-bright fast radio burst (FRB) from a modest, 3.4-day pilot survey with the Australian Square Kilometre Array Pathfinder. The survey was conducted in a wide-field  fly's-eye configuration using the phased-array-feed technology deployed on the array to instantaneously observe an effective area of $160$~deg$^2$, and achieve an exposure totaling $13200$\,deg$^2$\,hr . 
We constrain the position of FRB~170107 to a region $8'\times8'$  in size (90\% containment) and its fluence to be $58\pm6$~Jy~ms.
The spectrum of the burst shows a sharp cutoff above $1400$~MHz, which could be either due to scintillation or an intrinsic feature of the burst.
This confirms the existence of an ultra-bright ($>20$~Jy~ms) population of FRBs.

\end{abstract}


\section{Introduction} \label{sec:intro}

\begin{figure*}[!ht]

\includegraphics[width=0.9\linewidth]{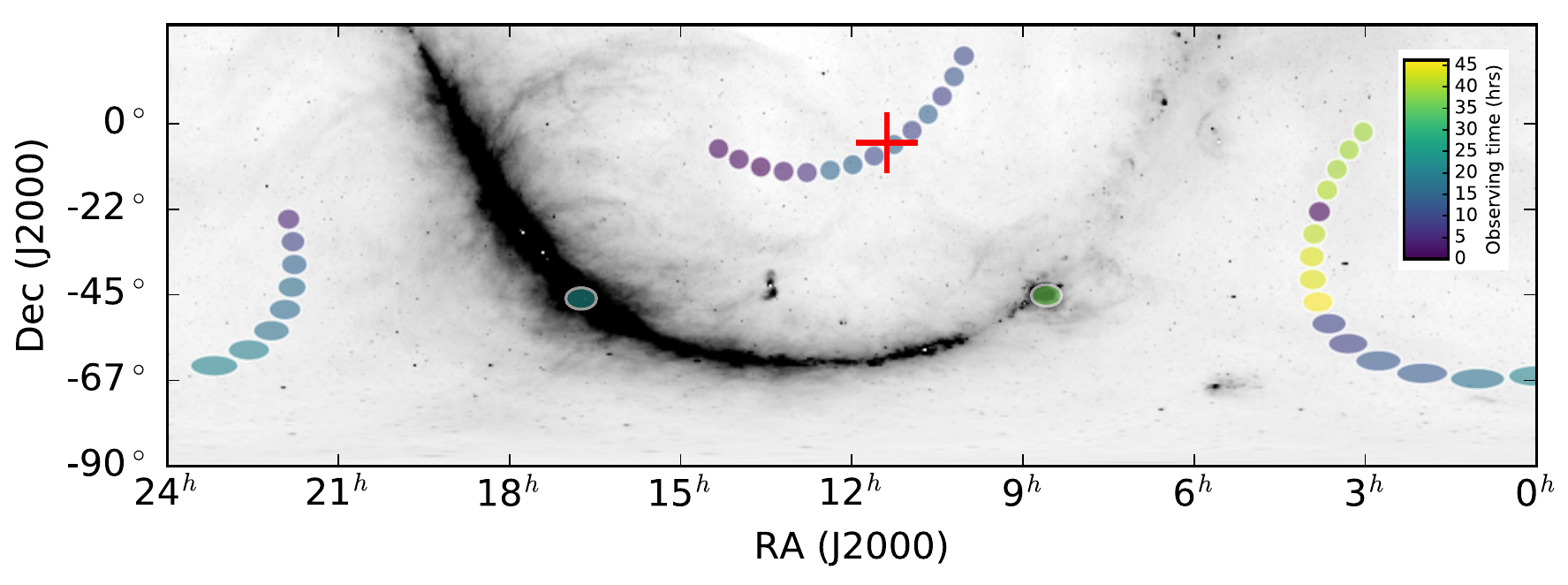}
\caption{\label{fig:pointings} CRAFT pointings overlaid on the CHIPASS \citep{Calabretta14} 1.4~GHz continuum map. The pointings are approximated as 5.5$^\circ$ radius circles and colored by number of observing hours.  The position of FRB~170107 is marked with a cross (+).  The two fields in the Galactic plane (RA, DEC) = ($8$\,h, $-45^\circ$) and ($16$\,h, $-45\circ$) are test fields in the directions of the pulsars B0833$-$45 and B1641$-$45. }
\end{figure*}

In the decade since the first reported detection of a fast radio burst \citep[FRB;][]{Lorimeretal2007}, { progress has been limited} in explaining the presence of this population of millisecond-duration radio pulses, which, based on their inferred extragalactic distances \citep{Thorntonetal2013}, are orders of magnitude more luminous than any pulse seen from the Milky Way.  
Despite a flurry of activity to detect more events \citep{Keithetal2010,Waythetal2011,Calebetal2016,Spitleretal2016,Masuietal2015,Chatterjeeetal2017} and an estimated rate of $\sim 3000\,$events\,sky$^{-1}$\,day$^{-1}$ above a fluence of $2$ Jy\,ms \citep{KeanePetroff2015,Championetal2016}, fewer than thirty such bursts are still known\footnote{see \url{http://www.astronomy.swin.edu.au/pulsar/frbcat/}} 
\citep{Petroffetal2016}. 

The ability to instantaneously survey the sky over as large an area as possible is crucial to detecting the rarest, most luminous bursts that are essential to understanding the highly problematic energetics of FRB radio emission.  The early recognition of a likely cosmological origin of these bursts, based on their high dispersion measures, led to energy estimates in the range $10^{31}$--$10^{33}\,$J \citep{Lorimeretal2007,Thorntonetal2013}.  All but two of the published fluence estimates are, in fact, lower limits, because the telescopes used to detect them undersample the focal plane, leading to large uncertainties of the burst positions within the beams.

This energetics problem has become more acute both with the confirmation that the bursts indeed originate at cosmological distances \citep{Tendulkaretal2017}, and with the recent detections of FRB~150807 and FRB~160317, extremely bright bursts which attained fluences of $\gtrsim 50$\,Jy\,ms \citep{Ravietal2016,Calebetal2017}.  Questions remain as to the prevalence of these bursts and how they relate to the rest of the FRB population.

A hitherto substantial impediment to the detection of FRBs has been the relatively small field of view available on currently operational telescopes.  
The Parkes radio telescope, equipped with a 13-beam multibeam receiver that has a $\approx 0.6$\,deg$^{2}$ field of view at 1.4\,GHz, has detected all but five of the FRBs published to date, with an average less than one event per 300~h.  Searches are often performed in conjunction with pulsar surveys \citep[e.g.][]{Keithetal2010}, which dedicate the majority of their time to searching for pulsars close to the Galactic plane.  This is problematic because the Parkes FRB detection rate appears to be a factor of 3--4 lower at low Galactic latitudes ($|b|<20^\circ$) than higher latitudes \citep[$|b| \gtrsim 50^\circ$,][]{Petroffetal2014}.

The Australian SKA Pathfinder \citep[ASKAP;][]{Johnston08,Schinckel12} is a next-generation wide-field telescope potentially capable of detecting FRBs at a rate an order of magnitude higher than previously possible.  
 This advance is enabled by the phased array feeds mounted on each of its 12-m antennas, yielding a $30\,$deg$^{2}$ field of view.  
The aim of the Commensal Real-time ASKAP Fast Transients \citep[CRAFT;][]{Macquartetal2010} survey is to equip ASKAP with high-time resolution capabilities that access its entire field of view on timescales of $1\,$ms.  

Here we report the first detection of an FRB with an array of ASKAP antennas operating in a fly's-eye configuration using the CRAFT observing mode and pipeline.

\section{Data capture and processing}
Each ASKAP antenna is equipped with a prime-focus phased array feed \citep[PAF;][]{HayOSullivan2008}, comprising 188 linearly-polarised receiving elements sensitive to frequencies between $0.7$ and $1.8$~GHz. The signal from each element is sampled at a input port and channelised to $1$~MHz frequency resolution. Digital beamformers form 36 dual polarisation beams by applying complex-valued weights to the individual ports, independently for each of $336$ channels, yielding a total bandwidth of $336$~MHz. In the observations presented here, the 36 beams are arranged in a $6\times6$ square pattern with a separation of $0.9^\circ$. The weights are calibrated using a maximum signal-to-noise (S/N) algorithm and the Sun as a reference source \citep[][and references therein]{Hotan14,McConnell16}. 
The observations were conducted in an eight-telescope fly's-eye mode, with each pointing in a different direction, yielding $8\times 36$ beams on the sky.

In the real-time CRAFT data pipeline, described in detail by \citet{Clarke14}, the beamformed voltages are squared and averaged over an integration time of $1.265$~ms and transmitted to a processing computer, which sums the polarizations and writes the data to disk.

We searched the data offline using a GPU-based FRB detection pipeline named the Fast Real-time Engine for Dedispersing Amplitudes (FREDDA; Bannister 2017, in prep.) that implemented the Fast Dispersion Measure Transform \citep[FDMT;][]{Zackayetal2017}.   FREDDA operates in blocks of 512 integrations and independently on each beam. First, FREDDA flags and rescales the incoming block. It then performs the FDMT of the flagged and rescaled data, computing 4096 dispersion trials with a DM resolution of $0.92$\,pc\,cm$^{-3}$ out to a maximum of $3763$\,pc\,cm$^{-3}$. Finally, FREDDA computes the boxcar convolution of each DM trial { with all widths in the range 1, 2, 3, ... 32 samples. }

Candidates exceeding $10 \sigma$ are written to disk and grouped offline using an iterative friends-of-friends algorithm similar to that developed by \citet{Huchra82}. { In the friends-of-friends algorithm each candidate is represented as a structure containing S/N, and a central DM and time. The structure also contains the extrema of all `friends' merged with the candidate in both time and DM. The algorithm is initialized with a list containing all detected candidates whose central and extreme times and DMs are set to the detected times and DMs, respectively. 

In an iteration the algorithm and finds all instances in the input list (`friends') within $32$ integrations and $20$ DM resolution elements of the extrema of each given candidate. It appends a new candidate to the output list whose extremities are set to the extrema of all the friends associated with the candidate. The S/N of the new candidate is set to the maximum S/N over all the friends, and the central time and DM to that corresponding to the friend with maximum S/N. The candidate and its friends are removed from the input list and the algorithm continues until the incoming list is exhausted. The algorithm terminates when no further candidates are merged.}

We plot for manual inspection those candidates whose properties match those consistent with extragalactic pulses, namely DM$>100$\,pc\,cm$^{-3}$. We further reject signals with widths exceeding $16$~samples, as their broad temporal signatures are characteristic of radio interference.

\section{Observations}
Observations were conducted between 2017 Jan $4$ and 2017 Jan $8$. We centered the observing band at $1.32$ GHz and observed regions of the  sky (see Fig. \ref{fig:pointings}) centered on Galactic latitudes of $|b| \approx 50~\deg$, sufficiently high to mitigate against possible latitude-dependent decrements in the detection rate and any bias that could be induced by observing over a range of latitudes \citep[see][]{Petroffetal2014,MacquartJohnston2015}. 

Hour-duration observations of the high-latitude fields were interleaved with 5-minute observations of the bright pulsars B0833$-$45 and B1641$-$45 to verify system performance. The pulsars were placed at the boresight of the antenna, equidistant from the central four beams.

The nominal system equivalent flux density (SEFD) for an individual beam from an individual antenna is $\approx 1800$~Jy. While the area enclosed by the half power contour of the 36-beam field is $\approx 30$\,deg$^2$, the sensitivity over the full field of view is nonuniform, and the equivalent performance (in survey speed terms) is of a uniform beam with an area of 20~deg$^2$ and a SEFD of $\approx 1800$~Jy \citep{Bunton2010}.  We adopt a field of view of $20\,$deg$^2$ in estimating the areal detection rate.  Interferometric measurements of the sensitivity (Table \ref{tab:frb_snr}) obtained on 2017 Jan 3 using the method of \citet{McConnell16} confirm the average SEFD of $\approx 2000$~Jy for all beams and all antennas used here.

\begin{deluxetable}{llrrr}
%\begin{center}
\tabletypesize{\scriptsize}
\tablecaption{FRB S/N and System performance\label{tab:frb_snr}}
\tablehead{Beam & $\Delta \alpha \cos \delta$  & $\Delta \delta$ & S/N  & SEFD\\
 & (arcmin) & (arcmin) & & (Jy)}
\startdata
13$^{\dagger}$ & 1.880 & 0.332 & 16.0 & 1800\\
12 & 1.143 & 0.848 & 4.6 & 1800\\
29 & 1.659 & 1.585 & 2.9 & 2100 \\
30 & 2.396 & 1.069 & 2.8 & 2200\\
31 & 3.134 & 0.553 & $-$0.1 &2400 \\
32 & 2.617 & $-$0.185 &3.7 &2100 \\
33 & 2.101 & $-$0.922 &0.3  & 2100 \\
14 & 1.364 & $-$0.406 &$-$0.7& 1800 \\
02 & 0.627 & 0.111  &1.1 & 1800 \\
\enddata
\tablecomments{The beam offsets in right ascension and declination, $\Delta \alpha \cos \delta$ and $\Delta \delta$ respectively, are measured relative to the boresight position of the antenna. ($\dagger$).}
\end{deluxetable}

\section{Results} \label{sec:Results}

We analyzed a total 660~antenna-hours of high-latitude pointings which detected 1273 candidates. 
We inspected these candidates first by plotting them in the time--DM plane.
We found that most were associated with internally generated RFI, and were clustered in time and DM on individual antennas;
approximately 20 candidates remained at this point.  
These candidates were individually inspected, and all but one were associated with signals which did not follow the cold plasma dispersion relation and were identified as interference.

\begin{figure}[!ht]
\begin{center}
\begin{tabular}{c}
\includegraphics[width=0.9\linewidth]{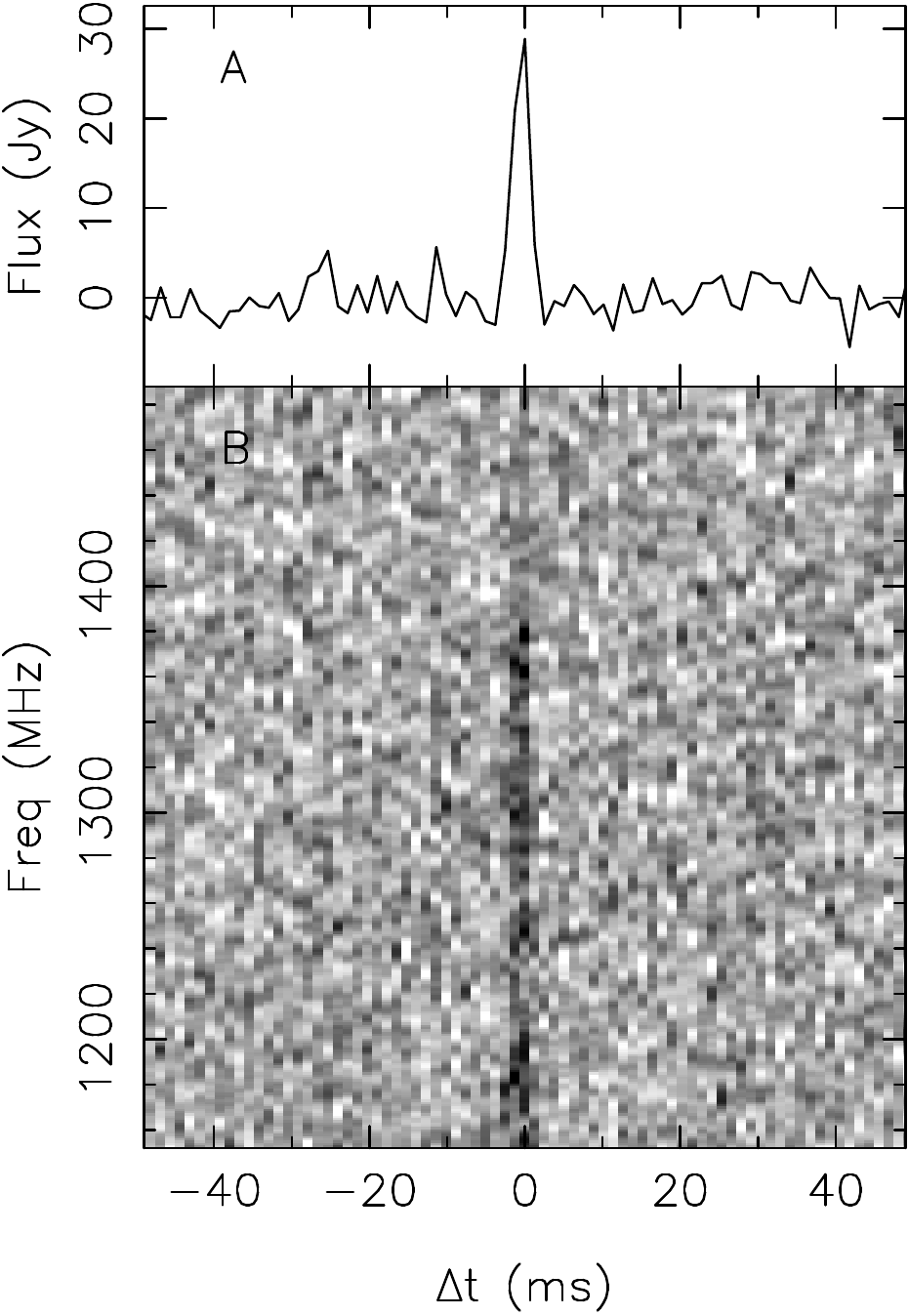} 
\end{tabular}
\end{center}
\caption{ FRB~170107 band-averaged pulse (panel A) and dynamic spectrum (panel B) from the data in beam 13 after dedispersion.
 \label{fig:frb170107}  }
\end{figure}

One candidate, FRB~170107, remained.  The pulse was detected in multiple beams; with the band-averaged pulse and its spectrum as detected in the primary beam are displayed in Figure \ref{fig:frb170107} and its key properties listed in Table \ref{tab:frb_properties}. 
The pulse dispersion measure, calculated by fitting the change in the pulse arrival time in the three equally spaced sub-bands below $1400\,$MHz, is $609.5\pm0.5$\,pc\,cm$^{-3}$, well in excess of the predicted contribution from our Galaxy \cite[see Table \ref{tab:frb_properties}, and][]{NE2001,YMW16}.
The pulse width is consistent with intra-channel dispersion smearing. 
There is no evidence for pulse broadening and we place a limit on the scattering time scale of $\tau_d \lesssim 2$\,ms. 

The pulse is detected in multiple beams of the PAF, as expected.  
In addition to the primary beam detection (beam 13), the pulse was detected with modest significance (S/N $\approx 3-4$) in two other beams (beams 12 and 32).   Other nearby beams have lower-significance detections and non-detections that are useful in constraining the location of the burst (and eliminating RFI as its origin, which would be present in all beams).  

We searched all observations of the field, totalling 15\,h, for other pulses at the same dispersion measure.  We found no candidates above a S/N ratio of 6.

\begin{deluxetable}{ll}
%\begin{center}
\tabletypesize{\scriptsize}
\tablecaption{Properties of FRB~170107\label{tab:frb_properties}}
\tablehead{} 
\startdata
Date   & 2017 Jan 07 \\
UTC  & 20:05:45.1397(1) \\
MJD & 57760.837328006(1)\\
Dispersion Measure (DM) &  609.5(5)~pc~cm$^{-3}$ \\ 
Right Ascension$^\dagger$ (J2000) &  $11^h 23 ^m 10^s (30)$   \\ %confirmed
Declination$\dagger$ (J2000) & $-05^{\circ} 01' (8)$  \\ %confirmed
Width & 2.6~ms \\
Fluence$^\dagger$   & 58(6) Jy~ms \\ \hline
\hline
Galactic latitude ($b$) & $51^\circ$  \\
Galactic longitude ($\ell$) & 	$266^\circ$ \\
\multicolumn{2}{c}{Assumed quantities} \\ \hline
DM$_{\rm MW, NE2001}$  & 35~pc~cm$^{-3}$\\
DM$_{\rm MW, YMW16}$ & 27~pc~cm$^{-3}$ \\
\enddata
\tablecomments{Uncertainties are $1-\sigma$ confidence regions, unless identified by dagger ($\dagger$); in that case they are $90\%$ confidence regions.  References:    NE2001: \cite[][]{NE2001}; YMW16:  \cite[][]{YMW16}.  } 
\end{deluxetable}

\subsection{Localization}

The multiple-beam detections enable the burst to be localized  to a region smaller than the beamwidth. This, in combination with knowledge of the beam shape, constrains the flux density, fluence, and spectral index of the burst.

We jointly model the detection in the primary beam with marginal detections and limits provided by the eight adjacent beams, using Bayesian methodology, in which we sample the posterior distribution: the product of a likelihood function and prior probability.  This allows us to robustly estimate parameters of interest and marginalize over other (nuisance) parameters.

We compare our measurements (or upper limits) on the pulse flux density in the primary beam detection and its eight adjacent beams to a model for the expected pulse flux density (expressed as a vector $\bm{S}_m(\bm{\theta},F)$)  that depends on its intrinsic brightness ($F$), and its location in the sky ($\bm{\theta}$). 
We assume that the measured flux densities ($\bm{\hat{S}}$)  have uncertainties that are normally distributed, but correlated.  

In this case, the logarithm of the likelihood function follows the usual form for Gaussian distributed noise:
\begin{equation}
-\log L = \left(\bm{\hat{S}} - \bm{S}_m(\bm{\theta}, F)\right)^T \bm{C}^{-1}  \left(\bm{\hat{S}} - \bm{S}_m(\bm{\theta}, F)\right),
\end{equation}
where $\bm{C}$ is the noise covariance matrix.    

As the PAF beams share common input ports, the noise in nearby beams is correlated, with adjacent beams having a correlation coefficient of $\rho\approx 0.2$, and the correlation decreasing in more distant beams \citep{Serraetal2015}.

To model the measured flux density of the burst $F$ and its position $\bm{\theta}$, is necessary to account for each beam's response to a signal  at arbitrary positions within the beam. 
We decompose the response into a relative beam sensitivity at boresight $g_i$, and a beam shape $B_i(\bm{\theta})$, that is unity at beam boresight. The beam sensitivity parameters $g_i$ are measured in contemporaneous interferometric observations of calibrators; we allow for fractional errors in them using the parameters $\sigma_{g,i}$.  
The measured flux density is therefore assumed to be
\begin{equation}
S_{i,m}(\bm{\theta}, F) =  F (1 + \sigma_{g,i}) g_i  B_i (\bm{\theta}).
\end{equation}

We assume a Gaussian shape for the beam model, parameterized by the  width $w_i(\nu)$ and location $\bm{\theta}_i$ of the beam.
The model is complicated by possible errors  in the beam position $\bm{\sigma}_{\bm{\theta},i}$ and width $\sigma_{w,i}$.   
In this case, the beam shape is
 \begin{equation}
B_i(\bm{\theta}) =  \exp\left[- \frac{\left( \bm{\theta} -\bm{\theta}_i - \bm{\sigma_{\theta,i}} \right)^2}{2 (1+\sigma_{w,i})^2 w_i^2(\nu)} \right],
\end{equation}
Holography of the antennas shows that the full width half power of the antenna beam is $w_{\rm FWHP}(\nu) =1.1c / (\nu D)\,$rad \citep{McConnell16}, where $D$ is the antenna diameter. As demonstrated below, the approximation of a Gaussian beam shape is sufficient to localize the burst position within the beam given its S/N.

We assume a uniform prior on $\bm{\theta}$ and a logarithmic prior on $F$.
Choice of a logarithmic prior on $F$ is typical for positive-definite quantities that have unknown value. 
We note however that this choice does (compared to i.e. a uniform prior in $F$) does not significantly affect parameter estimation because the flux densities are well constrained. 
To model potential errors in  beam sensitivities and shapes, for each beam, we apply a Gaussian prior on the beam-dependent width and gain variations (respectively $\sigma_{w,i}$ and $\sigma_{g,i}$), with both having a width of 0.1 (corresponding to 10\% root-mean-square variations). 
We assume that the beam positions can potentially be in eror by 1~arcmin (rms) in two independent orthogonal components, so also apply a Gaussian prior on $\bm{\sigma_{\theta,i}}$ of width $1$~arcmin.

We used the Multinest algorithm  \cite[][]{2009MNRAS.398.1601F} to  sample the posterior distribution. 

\begin{figure*}
\begin{tabular}{ccc}
\hspace{-1cm}
\includegraphics[width=0.3\linewidth]{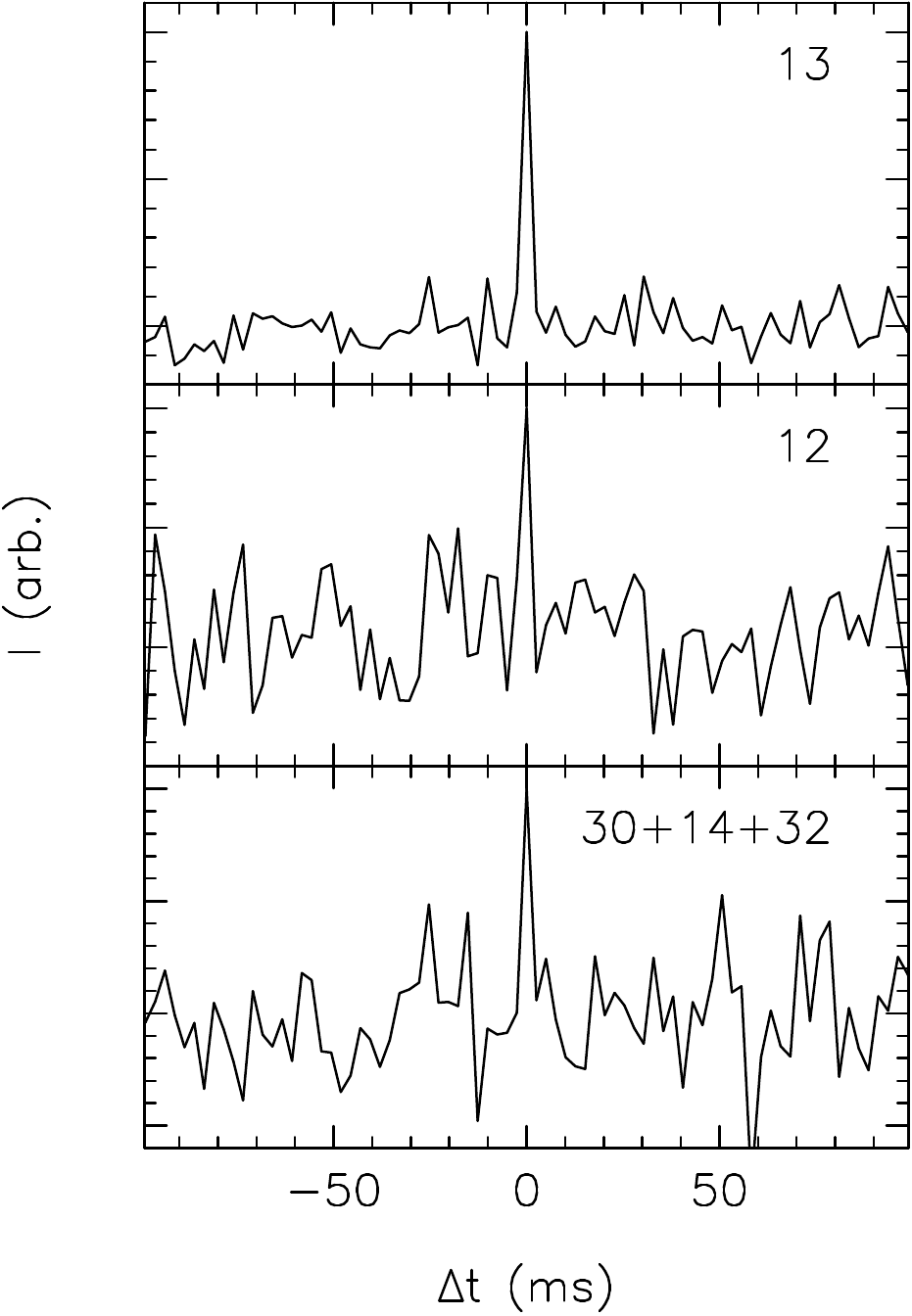}
\includegraphics[width=0.3\linewidth]{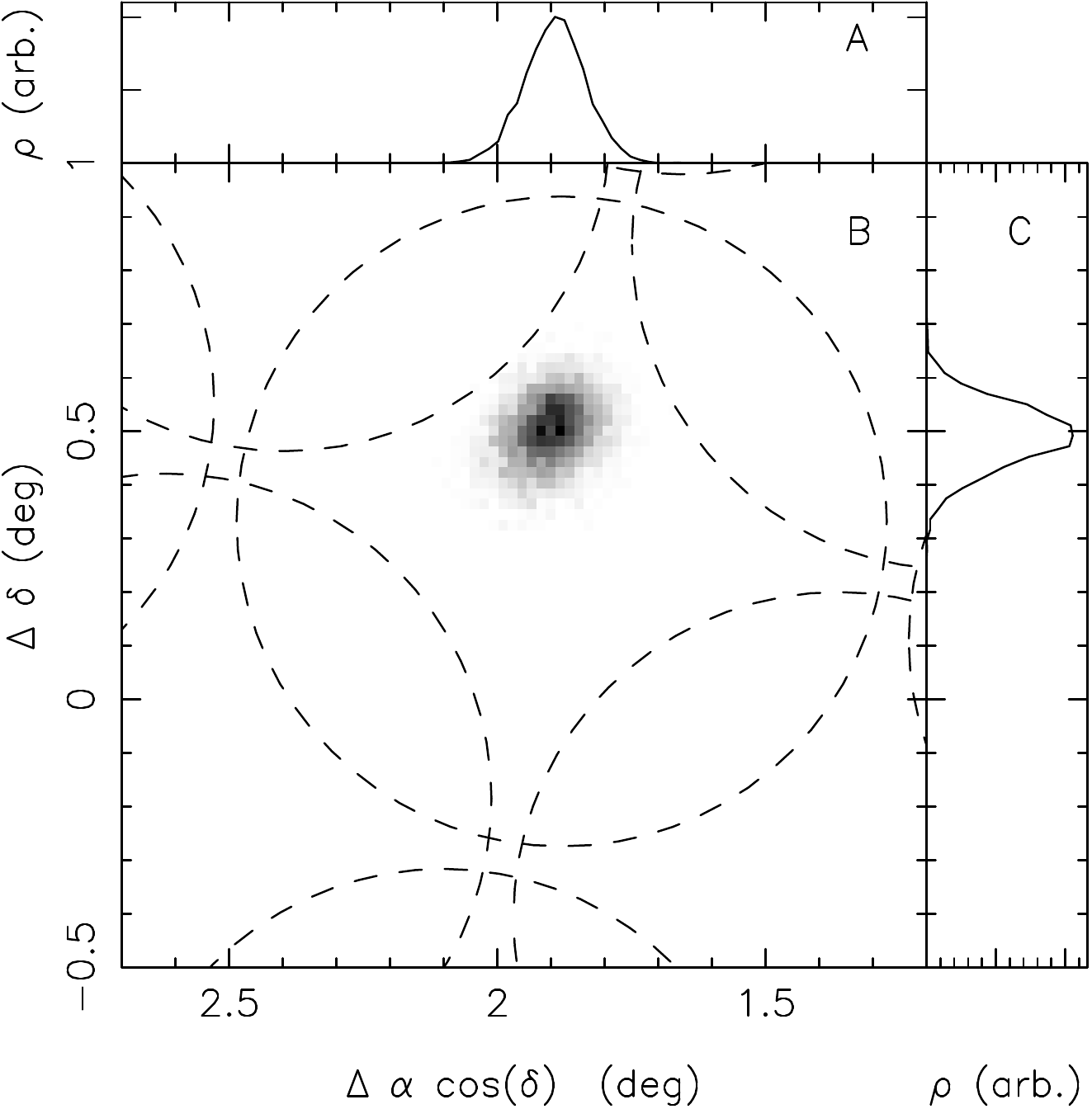}
\includegraphics[width=0.5\linewidth]{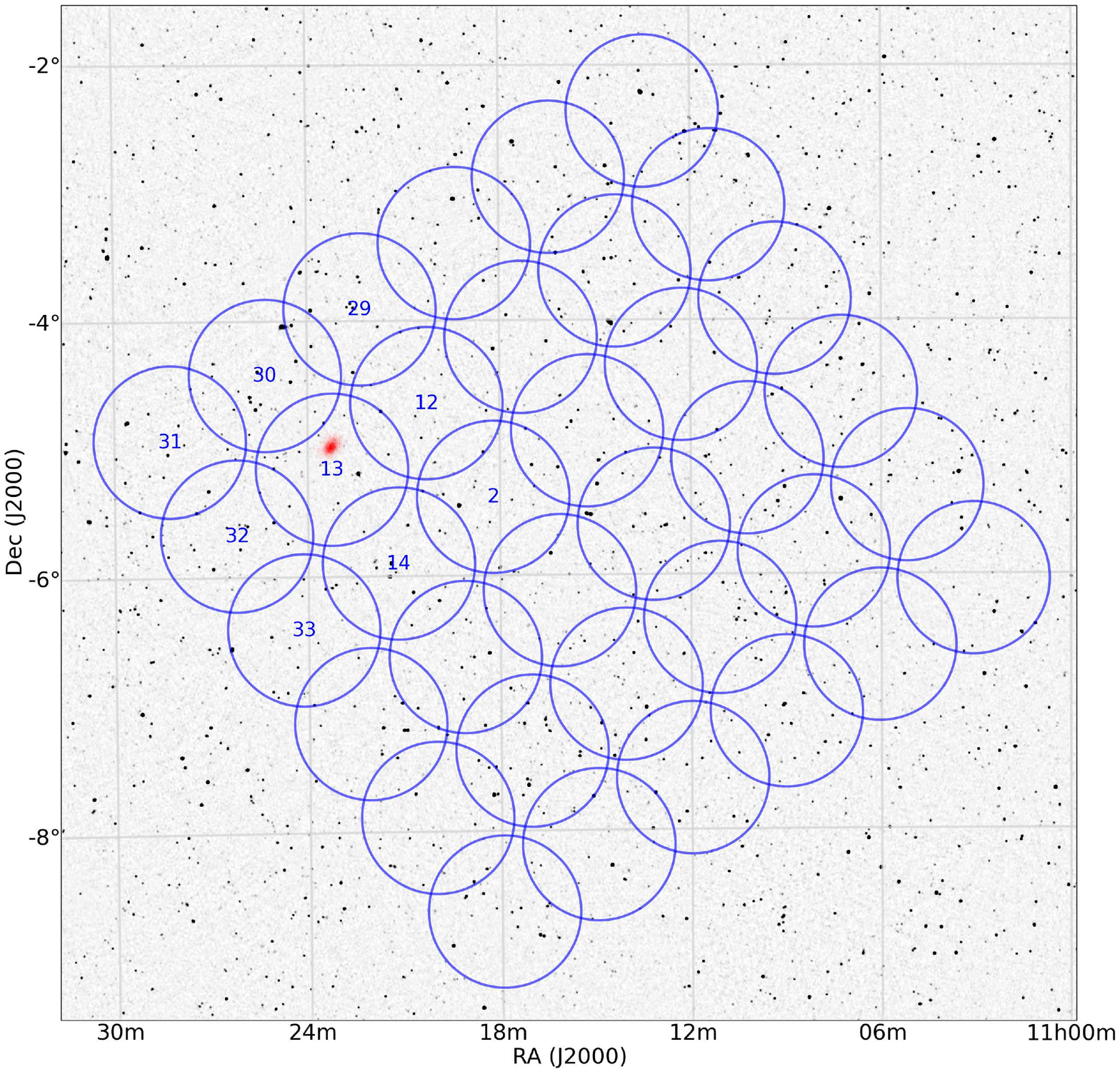}
\end{tabular}
\caption{\label{fig:localisation2} Left:  Multi-beam detections of FRB~170107, from beam 13 (top); beam 12 (middle); and the sums of beams 30, 14, and 32.   Right: Localization of FRB~170107.  The region of significant ($>99.99\%$) posterior probability density for the position of the FRB is shaded in red.  The assumed half power beam contours (1.2$^\circ$ at 1.3~GHz) of all the beams are shown in blue.  The background grayscale is from the the NRAO-VLA Sky Survey \cite[NVSS,][]{nvss}.  }
\end{figure*}

The modeling constrains the burst to a region of $8'\times8'$ in size at the 90\% confidence level and $2'\times2'$ at the 50\% confidence level.  In Figure \ref{fig:localisation2}, we show the burst localization region relative to the nearby beams (centre panel) and to the field of view of antenna AK05 (right panel).

\begin{figure}[!ht]
\begin{center}
\begin{tabular}{c}
  \includegraphics[scale=0.5]{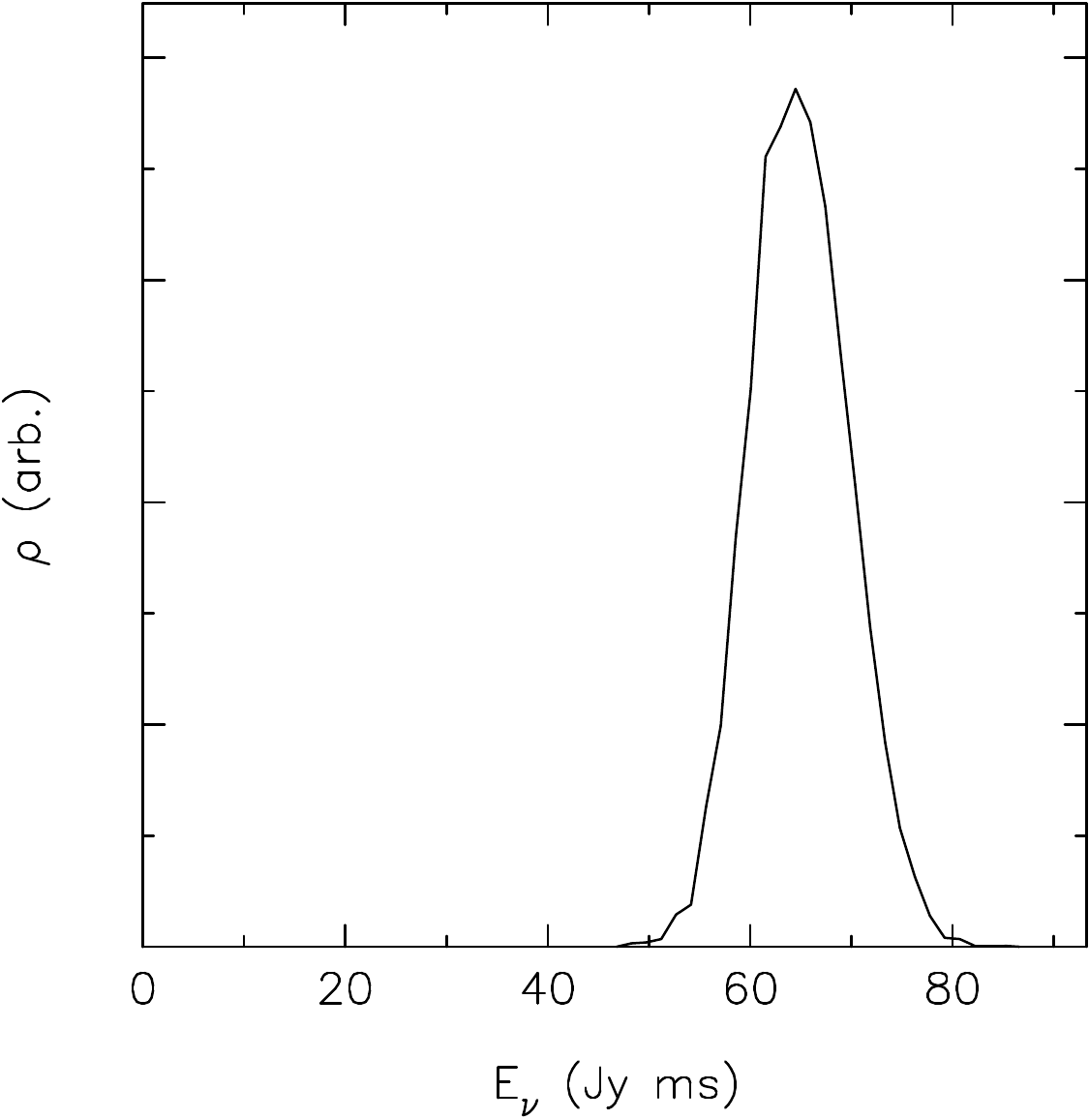}
\end{tabular}
\end{center}
\caption{Posterior fluence distribution of FRB~170107. 
 \label{fig:posterior_energy}  }
\end{figure}

As a consequence of the tight localization relative to the primary beam, we are able to constrain the pulse peak flux density to be $27 \pm 4$~Jy and its fluence to be $58\pm6$~Jy~ms (90\% confidence; see Figure~\ref{fig:posterior_energy}).

As the pulse was detected close to the center of beam 13, the spectral variation of the pulse is not an instrumental effect.  Below the cutoff at 1400\,MHz, we find the spectral index of the burst $\alpha=-0.4\pm0.8$ ($S \propto \nu^{\alpha}$).

 We confirmed our methodology by localizing the bright pulsar B1641$-$45 to $\approx 0.6'$, using an observation taken with antenna AK05 approximately $30$\,minutes  after the FRB detection.

\subsection{UTMOST followup}

The refurbished Molonglo Radio Telescope (UTMOST; Bailes et al.\,2017, in prep.) followed up the FRB field for a total of 26.1\,h in March
2017 -- typically between 3 and 6\,h in each observation.
The telescope operates at a central frequency of 834\,MHz and has an
effective 16\,MHz of bandwidth, with a single circular polarization.
The observations were conducted using a mode that produces 352 fan-beams of size
$46''\times2.8^\circ$, tiled across the $4.0^\circ\times2.8^\circ$ primary
beam. 

The UTMOST frequency channels are 98\,kHz wide, leading to a DM smearing of 0.8\,ms at the DM of FRB~170107. 
No candidate events were found to match a repeat burst from FRB~170107. 
The search depth to S/N = 9 corresponds
to a fluence limit for no repeat bursts for the FRB to $\approx 30
\times (W/1\,\mathrm{ms})^{0.5}$ Jy\,ms (at the telescope boresight),
where $W$ is the pulse width.

\section{Discussion} \label{sec:Discussion}

The detection of FRB~170107 highlights the capacity of ASKAP and future widefield telescopes to access fundamentally new regions of transients parameter space.  
The large instantaneous field of view of the array, a factor of 400 greater than the Parkes radio telescope (using the 13-beam multibeam receiver),
has resulted in the detection of a rare but bright FRB.   

By taking advantage of the dense (Nyquist) sampling of the focal plane enabled by the PAFs, we have constrained several burst properties more tightly than has hitherto been possible with single-dish telescopes.  Specifically we have localized the burst to a region a factor of $60$ smaller than the beam area, and thus robustly determined its fluence and spectrum.

Our localization is comparable in size to the best obtained with a non-interferometric instrument, that of FRB~150807, detected with the much larger 64\,m Parkes telescope. 
The localization is insufficient to identify a unique host;  however, there are no cataloged bright galaxies in the field.  This is consistent with the burst originating at cosmological distances or in a smaller, possibly closer, uncatalogued galaxy.

As the burst location is close to the center of the primary beam, we are confident that the observed spectral modulation of the burst, in particular the strong decrease in flux above $1400$~MHz, is not an instrumental effect.  This spectral feature is either intrinsic to the burst or the result of diffractive scintillation, due either to the ionized interstellar medium of the Milky Way or the burst host galaxy.  High-latitude lines of sight in the Milky Way typically have scintillation bandwidths of tens to hundreds of MHz, comparable to the width implied by the spectral modulation observed here. Diffractive scintillation has also been posited to present an observational bias that boosts the intrinsic flux densities of some FRBs, and introduces the observed latitude dependence of the FRB rate \citep{MacquartJohnston2015}.   

Similar dropouts are seen in other FRBs, notably the repeating FRB \cite[FRB121102;][]{scholz16}, and it is similarly unclear whether these are intrinsic to the bursts or the result of scintillation.

%Redshift
We can  estimate a distance to the burst under the assumption that the bulk of its dispersion measure is due to the ionized Inter-Galactic Medium (IGM).  Using the estimated Galactic contribution to the DM along the light of sight to FRB~170107 of $35\,$pc\,cm$^{-3}$ (see Table \ref{tab:frb_properties}), and assuming a host contribution of a further $100\,$pc\,cm$^{-3}$, we estimate the burst redshift\footnote{Throughout this text we assume $\Omega_m$=0.27, $\Omega_\Lambda$=0.73, $H_0$=70\,km\,s$^{-1}$\,Mpc$^{-1}$, and a baryon density at $z$=0 of $1.8 \times 10^{-7}$\,cm$^{-3}$.} to be $z\approx 0.54$.  However, this estimate is also subject to the assumption that the IGM is smoothly distributed along the line of sight; random contributions by halos add considerable uncertainty to this value \citep[see][]{McQuinn2014}.  At this nominal redshift, this luminosity distance is $3.1$\,Gpc, and the total estimated isotropic-equivalent burst energy across the observing band is $\sim 3\times 10^{34}\,$J.

Our result demonstrates that a population of bright FRBs exists.  Six of the $23$ published bursts have fluences $> 20$\,Jy\,ms, a population easily probed by ASKAP.  {

Our implied detection rate is broadly consistent with that predicted by \cite{KeanePetroff2015}, who estimated there are $2.5_{-1.6}^{+3.2} \times 10^3$\,events sky$^{-1}$\,day$^{-1}$ for fluences $> 2\,$Jy\,ms.  The completeness limit of our survey is not yet precisely constrained, but a preliminary estimate of 20\,Jy\,ms implies that the survey should have detected $3.3_{-1.1}^{+4.3}$, $1.1_{-0.7}^{+1.3}$ or $0.3_{-0.2}^{+0.4}$ events for a cumulative fluence count distribution, $N(>E_\nu) \propto E_\nu^{\alpha}$ with $\alpha$ respectively $-1$, $-1.5$ (i.e. Euclidean) or $-2$.  
However, our rate is subject to considerable additional uncertainty: compared to other 20-cm surveys, our fluence completeness suffers from dispersion smearing at lower DMs because of our worse spectral resolution.  We defer a definitive statement on the fluence distribution until ASKAP has acquired a large, controlled burst dataset.}

The direct, arcsecond localization of FRBs will be necessary to identify host galaxies, and to determine if they reside in radio nebulae and dwarf galaxies, like the repeating FRB~121102 \citep{Chatterjeeetal2017}. { The connection between repeating and the apparently non-repeating FRBs is the  major open observational question in FRB astrophysics and, with its wide field of view, and added high time resolution interferometric capabilities, ASKAP is poised to resolve it.}

\acknowledgments

The Australian SKA Pathfinder is part of the Australia Telescope National Facility which is managed by CSIRO. Operation of ASKAP is funded by the Australian Government with support from the National Collaborative Research Infrastructure Strategy. ASKAP uses the resources of the Pawsey Supercomputing Centre. Establishment of ASKAP, the Murchison Radio-astronomy Observatory and the Pawsey Supercomputing Centre are initiatives of the Australian Government, with support from the Government of Western Australia and the Science and Industry Endowment Fund. We acknowledge the Wajarri Yamatji people as the traditional owners of the Observatory site.  SO acknowledges Australian Research Council grant Laureate Fellowship FL150100148. This research is supported by the Australian Research Council through grants CE110001020 and FT150100415. We would like to thank the MWA director, Randall Wayth, for giving us access to the Galaxy GPU cluster.

\bibliographystyle{aasjournal}

\end{document}